# Engineering of frustration in colloidal artificial ices realized on microfeatured grooved lattices

Antonio Ortiz-Ambriz[1,2] & Pietro Tierno[1,2]

Artificial spin ice systems, namely lattices of interacting single domain ferromagnetic islands, have been used to date as microscopic models of frustration induced by lattice topology, allowing for the direct visualization of spin arrangements and textures. However, the engineering of frustrated ice states in which individual spins can be manipulated *in situ* and the real-time observation of their collective dynamics remain both challenging tasks. Inspired by recent theoretical advances, here we realize a colloidal version of an artificial spin ice system using interacting polarizable particles confined to lattices of bistable gravitational traps. We show quantitatively that ice-selection rules emerge in this frustrated soft matter system by tuning the strength of the pair interactions between the microscopic units. Via independent control of particle positioning and dipolar coupling, we introduce monopole-like defects and strings and use loops with defined chirality as an elementary unit to store binary information.

[1] Department of Structure and Constituents of Matter, University of Barcelona, Avinguda Diagonal 647, 08028 Barcelona, Spain. [2] Institute of Nanoscience and Nanotechnology, IN2UB, Universitat de Barcelona, 08028 Barcelona, Spain. Correspondence and requests for materials should be addressed to P.T. (email: ptierno@ub.edu).

Geometric frustration emerges in disparate physical and biological systems that span from disordered solids[1], to trapped ions[2], ferroelectrics[3], microgel particles[4], high-$T_c$ superconductors[5], folding proteins[6] and neural networks[7]. When topological constraints between the individual elements impede the simultaneous satisfaction of all local interaction energies, the system is geometrically frustrated, featuring a low-temperature residual entropy and a large ground state (GS) degeneracy[8], as observed in water ice[9] and in rare-earth pyrochlore oxides, called spin ice[10–12]. On the theoretical side, topologically frustrated spin systems have been scrutinized for a long time, dating back to the work of Wannier[13] on the Ising model applied to a triangular lattice, in which the system cannot accommodate three spins on each plaquette in such a way that all antiferromagnetic couplings are minimized. In pyrochlore crystals, the situation is similar: the rare-earth ions carry a net magnetic moment, and they are located on the sites of a lattice of corner-sharing tetrahedra[10]. At each vertex, the moments can point either towards the centre of the tetrahedron or away from it, and pairs of spins align in the low-energy head-to-tail configuration. The degenerate GS of pyrochlore crystals follows the ice-rules[14] where two spins point towards the centre of the tetrahedron and two away from it. The ice rules were first introduced by Bernal and Fowler[14] to describe the proton ordering in water ice (ice $I_h$). In the hexagonal $I_h$, the lowest energy configuration is characterized by two protons near an oxygen ion in a tetrahedron and two away from it similar to the spin ice systems. To fulfil these rules, there are six equivalent atomic configuration at each tetrahedron and Pauling showed that this degeneracy was at the origin of the residual entropy of water at low temperature[15]. However, the ice rules can be locally violated[16] due to the presence of disorder or fluctuations in the system, giving rise to mobile excitations that mimic the behaviour of magnetic monopoles and Dirac strings[17–20].

Investigating the governing rule in frustrated systems is a key issue not only for understanding exotic phases in magnetism but also for providing guidelines to engineer new magnetic memory and data processing devices[21]. However, experimental investigations of bulk spin ice materials have often been restricted to averaged quantities, such as heat capacity curves[22], magnetic susceptibility[23] or neutron scattering data[24]. Artificial spin ice has recently been introduced as an alternative system that displays ice-like behaviour and allows for the direct visualization of the individual spins[25]. Such systems are composed of lithographically fabricated ferromagnetic nanostructures[25,26] or nanowires[27] arranged into periodic lattices that generate frustration by design. Given the experimentally accessible length-scale of a few nanometres, the spin orientation and the system GS can be visualized by using magnetic force microscopy, although monitoring the dynamics leading to the system degeneracy remains an elusive task because of the extremely fast spin flipping process.

Here we engineer a mesoscopic artificial spin ice that consists of an ensemble of interacting colloidal particles confined to lattices of gravitational traps. Our experimental system is inspired by a recent theoretical proposition[28], in which electrically charged colloids in a square lattice of bistable optical traps were observed to obey ice-rule ordering for strong electrostatic interactions. The high demand in laser power required to generate the necessary optical traps combined with the difficulty of tuning the surface charge in colloidal systems motivates the use of an alternative approach. We overcome these problems by using interacting microscopic magnetic particles confined to the lattices of bistable gravitational traps. The colloidal spin ice allows us to probe the equilibrium states and the dynamics of pre-designed frustrated lattices, and provides guidelines to engineer novel magnetic storage devices based on frustrated spin states.

## Results

**Realization of the colloidal spin ice.** We use paramagnetic particles with tunable interactions inside lithographically sculptured double-well traps, as shown in Fig. 1a. Each trap is fabricated by etching an elliptical indentation in a photocurable resin and leaving a small hill in the middle. We arrange these bistable traps into honeycomb or square lattices, although different lattice conformations can be easily implemented by lithographic design. The elliptical wells have an average length of 21 µm, width of 11 µm and we use a lattice constant of $a = 33$ µm for the honeycomb lattice and $a = 44$ µm for the square lattice. Paramagnetic microspheres of diameter $d = 10.4$ µm are dispersed in water and then allowed to sediment above the surface of the resin. Later, the particles are placed in the traps at a one-to-one filling ratio using optical tweezers, (see Methods section). Within the double wells, the colloidal particles are gravitationally trapped in one of the two low-energy states. A typical profile of the double wells obtained via an optical profilometer is shown in Fig. 1b. With this technique we measure an average barrier height within the well of $h = 0.43 \pm 0.04$ µm, where Fig. 1c shows a typical well of $\sim 0.3$ µm height. Given that the density mismatch is $\Delta\rho = 0.9$ g cm$^{-3}$ between the particles and the suspending medium, we estimate a gravitational energy in the centre of the bistable trap of $U_g = 540 k_B T$, and an outer confining potential for each island of $\sim 3,000 k_B T$, where $k_B$ is the Boltzmann constant and $T = 20$ °C is the experimental temperature. Thermal fluctuations are unable to induce spontaneous switching of the particle state unless either smaller particles or a smaller hill are used.

To tune the pair interaction between the magnetic colloids, the paramagnetic particles are doped with nanoscale iron oxide grains; as a result of doping, these particles are responsive to a magnetic field **B**. Under the applied field, the particles acquire a dipole moment $\mathbf{m} = V\chi\mathbf{B}/\mu_0$, where $V = (\pi d^3/6)$ is the particle volume, $\chi = 0.1$ the magnetic volume susceptibility and $\mu_0 = 4\pi 10^{-7}$ H m the susceptibility of vacuum. Pairs of particles interact via dipolar forces, and for a magnetic field applied perpendicular to the plane, the interaction potential is isotropic and is given by $U_d = \frac{\mu_0 m^2}{4\pi r_{ij}^3}$ where $r_{ij} = |\mathbf{r}_i - \mathbf{r}_j|$, $r_i$ is the position of particle $i$. We apply a homogeneous field ranging from 0 to 25 mT with an accuracy of 0.1 mT. When the paramagnetic colloids cross the central hill, we find that the corresponding out-of-plane motion produces a negligible effect on the overall collective dynamics.

**Spin configuration and vertex energy.** Two typical experimental realizations are shown in Fig. 1d,e for honeycomb and square lattices, respectively, both following exposure to a constant field of amplitude $B = 18$ mT for 60 s. Each experiment is initialized by loading one particle in each well using optical tweezers, and randomly flipping the position within the well according to a random number generator. The repulsive magnetic interactions force the particles to maximize their distance and are such that the particles can easily switch state but cannot escape from the gravitational trap. By assigning a vector (analogous to a spin) to each bistable trap pointing from the vacant site towards the side occupied by the particle, one can construct a set of ice rules for the colloidal spin ice system, equivalent to those for artificial spin ice[28]. At each vertex where three traps (in the case of honeycomb) or four traps (in the case of square) meet, the vector assigned to each trap can point either in when the colloidal particle is close to the vertex or out when it is far from the vertex, following the same classification scheme used in the three-dimensional pyrochlore tetrahedron. The vertices of the honeycomb lattice, sometimes called kagome ice since the spin midpoints are arranged in a kagome lattice, can have four different types of spin arrangements. The highest energy configuration occurs when three particles are close to the vertex ($K_{IV}$), and the lowest energy configuration has three particles far from the vertex ($K_I$). In contrast, the square lattice has six types of spin configurations: the highest energy vertex is composed of four particles close to each other ($S_{VI}$), and the lowest energy vertex has all the particles far away ($S_I$). The corresponding energetic weight of all the vertices in both lattices is shown at the bottom of Fig. 1. In particular, for the elliptical wells used in Fig. 1, the magnetic interaction between nearest neighbours can vary from $U_d = 6425\,k_BT$ to $U_d = 450\,k_BT$ in the honeycomb lattice and from $U_d = 1675\,k_BT$ to $U_d = 228\,k_BT$ in the square lattice. These interactions increase as more particles are added at each vertex. Moreover, we notice that in contrast to the artificial spin ice, the colloidal system is characterized by mobile particles and their pair interaction depends on the relative distance between them. The energy at each vertex thus changes as the particles move, and the corresponding GS results from a collective effect between the interacting units.

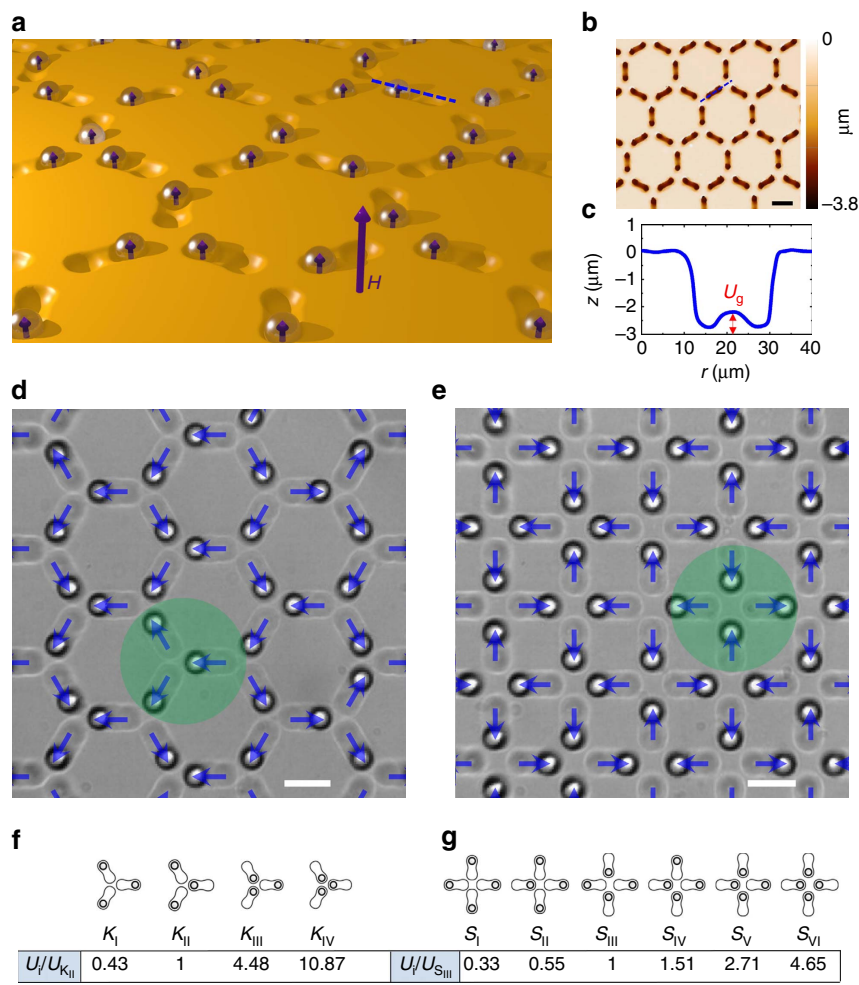

**Figure 1 | Realization of the colloidal spin ice system.** (**a**) Schematic view of the colloidal spin ice made by a honeycomb lattice of double-well islands filled with paramagnetic colloids. The applied field **B** perpendicular to the plane induces repulsive dipolar interactions between the particles. (**b**) Optical profilometer image of the honeycomb spin ice, and (**c**) the cross-section of a double well with a small central hill, giving a gravitational potential $U_g$. (**d,e**) Equilibrium state of a honeycomb ice (**d**) (lattice constant $a = 44\,\mu m$) and a square ice (**e**) (lattice constant $a = 33\,\mu m$). Blue arrows denote spin direction, while green circles highlight vertices of type $K_{II}$ (in **d**) and $S_{III}$ (in **e**). Scale bars, 20 μm for all images. (**f,g**) Vertex configurations for honeycomb (**f**) and square (**g**) ices. The lowest panel shows the normalized magnetostatic energy for each type of vertex.

**Ice-selection rules in the colloidal spin ice.** Systematic experiments performed by increasing the interaction strength via the applied field reveal that the colloidal spin ice has a clear tendency to follow ice-selection rules for both the square and honeycomb configurations. Figure 2a,b shows that the fraction of low-energy vertices of type $S_{III}$ and $K_{II}$ increases up to $\sim 0.8$. We confirm both trends by performing Brownian dynamic simulations, shown as continuous lines in Fig. 2a,b. In these simulations, we neglect many-body effects due to the relatively large separation between the interacting particles at each vertex, more details can be found in the Methods. In the ferromagnetic artificial square ice, Wang et al.[25] found that as the interaction increases, the system is dominated by vertices of types $S_{III}$ and $S_{IV}$. In contrast, we observe that when the applied field increases, the $S_{III}$ vertices dominate over the $S_{IV}$ vertices due to their slightly lower energetic weight. This is closer to the true GS of the square ice, which corresponds to a lattice that is fully covered by $S_{III}$ vertices. To obtain the GS in the magnetic bar system, Zhang et al.[26] recently used a dedicated annealing procedure based on a rotating demagnetizing field, while for the colloidal spin ice system a long-range ordered GS arises by simply increasing the interactions strength between the magnetic particles. In the case of the honeycomb lattice, a different set of ice rules arises, in which the high-energy $K_{IV}$ vertices and their topologically connected $K_I$ disappear in favour of the $K_{II}$ and $K_{III}$ vertices. In the colloidal system, beyond a field of $B = 9$ mT, the $K_{II}$ vertices begin to prevail, because they are energetically more favourable and maximize the average particle distance. In reality, at much higher strength, the number of observed $K_{II}$ and $K_{III}$ vertices should converge due to the isotropy of the honeycomb lattice vertices compared with the square one.

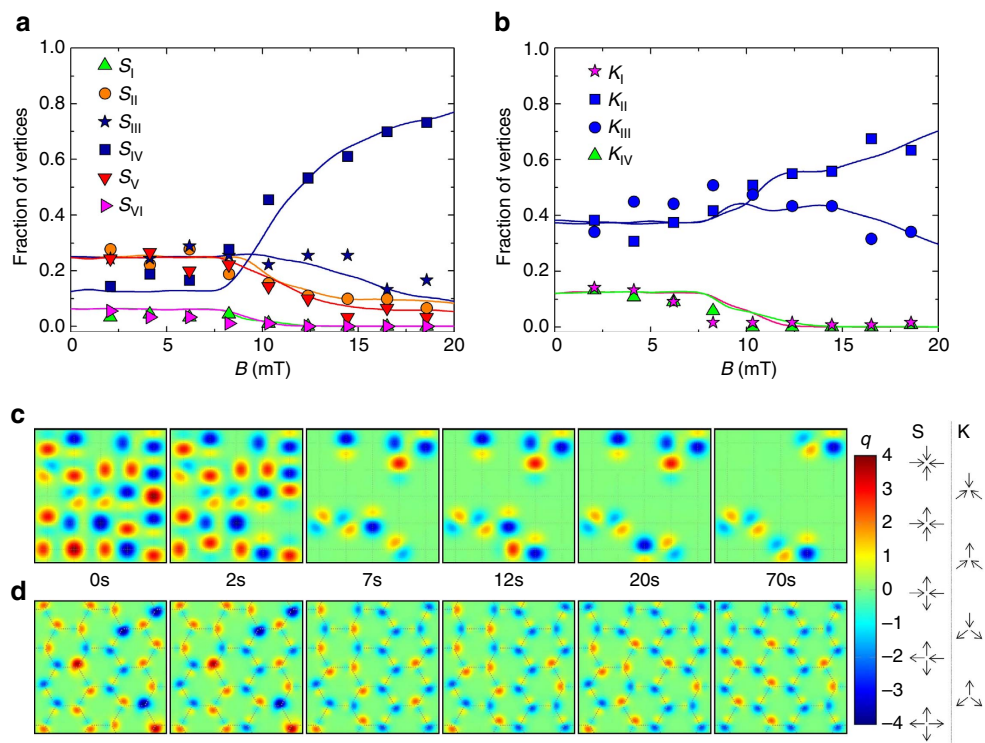

**Figure 2 | Ice rules and dynamics in colloidal spin ice.** (**a,b**) Average fraction of vertices at equilibrium for square (**a**) and honeycomb (**b**) ice versus the applied magnetic field. In both cases, ice-selection rules (blue) emerge for large interaction strengths. Scattered points are experimental data, continuous lines are results from Brownian dynamic simulation (Methods). The experimental points were averaged over 10 realizations, the numerical data over 1,000. (**c,d**) Evolution of the net vertex charge $q$ in the square (**c**) and honeycomb (**d**) ice (experimental system) after the application of an external field with amplitude $B = 18.5$ mT. Point charges at each vertex were assigned by considering a positive value for an incoming spin and a negative value for an outgoing spin, relative to the centre of the vertex. The schematics on the side of the colour bar illustrate the total sum of the charges $q = 0; \pm 2; \pm 4$ for the square ice and $q = \pm 1; \pm 3$ for the honeycomb ice.

**Particle dynamics above the square and honeycomb lattices.** One advantage of our mesoscopic system is that by using particle tracking routines, we can directly follow the colloid displacement and monitor the entire ordering process for a given interaction strength. The dynamics are better visualized by displaying the net topological charge $q$ at each vertex, calculated according to the dumbbell model, in which each spin carries a pair of opposite charges at each of its ends[25]. Colour maps of the charge in the system at different times are shown in Fig. 2c,d for the square and honeycomb lattices, respectively. The GS for the square ice corresponds to vertices with a zero net charge. The highly frustrated honeycomb ice shows how low-energy vertices with net charge $q = \pm 1$ are preferred over high-energy vertices with $q = \pm 3$. Starting from an initial random distribution of particles within the traps, the square ice shifts towards a state without highly charged defects. However, some defects do remain frozen close to the sample edges, and the system converges to a low-energy metastable state. In contrast to the square ice, the honeycomb spin ice has two equivalent vertices and thus an inherent extensive degeneracy. It has been predicted to undergo a series of phases (Ising paramagnet, Ice I, Ice II and 'solid' ice) as the temperature decreases[27]. In Fig. 2d, we observe that the system organizes into a superlattice region of $+1$ and $-1$ charged vertices when a strong magnetic field is applied. This is more similar to the Ice II phase with the presence of few defects located at the edges which break the long-range order present in the solid ice phase. We note that the order observed in both types of lattices can be further improved either by applying annealing protocols with dynamic fields or by applying a bias force obtained by a strong magnetic gradient[29].

**Discussion**
Our system allows us to manipulate individual particles within the wells using optical tweezers. This method can be used to introduce defects, which can be later erased by turning on the magnetic field and thus increasing the interaction between the particles. We demonstrate this feature with the square lattice, although we have the same degree of control over the honeycomb ice. For the colloidal square ice, the spin directions in the GS (Fig. 3c) define chiral cells, which alternate in chirality in a checkerboard pattern. Pairs of defects emerge in the form of achiral cells with a net excess of magnetic charge when flipping one spin from the GS configuration. Depending on the location of these cells within the array, defects can be made stable or unstable at a given applied field. In Fig. 3a–c, we observe the evolution of a pair of defects with opposite charges, separated by a string of achiral domains indicated by the green crosses in Fig. 3a. When the annihilation of these defects involves only a few spin flips, the energetic cost for the system to recover its GS is low, and the defects rapidly disappear as the field is turned on, as shown in the sequence in Fig. 3b. The defect annihilation process occurs via a stepwise flipping of the particle position rather than via a cooperative shift of all spins simultaneously, as seen in Fig. 3f. In contrast, in Fig. 3d,e, where we show a string of defects that acts as a domain wall separating two GS regions with unmatched chirality, defective vertices are more difficult to erase because they require an entire region to flip to escape this metastable state. As a result, the domain wall remains practically frozen in place. The presence of large GS regions separated by domain walls emerges as a natural low-energy metastable state in ferromagnetic spin ice[19], even after a subsequent thermally induced annealing process[30], given the elusive nature of the true long-range ordered GS.

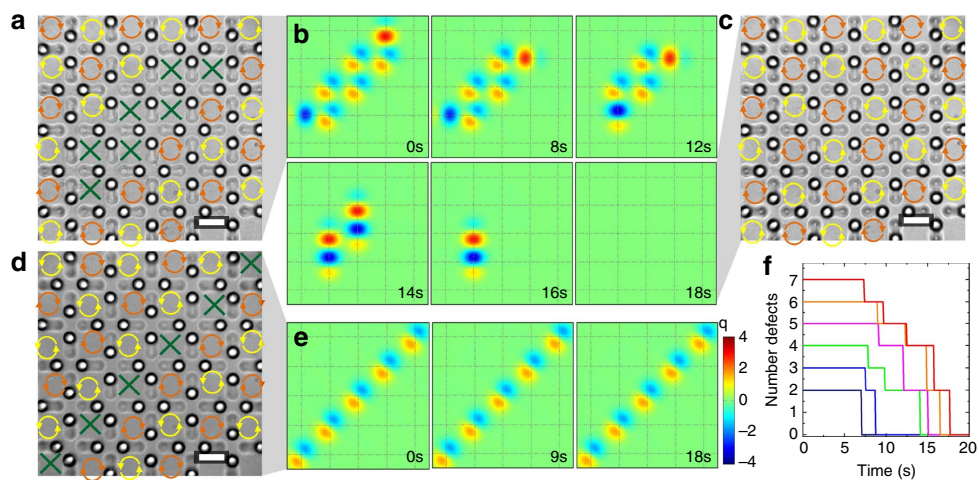

**Figure 3 | Defect annihilation and dynamics.** (**a**) Microscope image showing a square ice state prepared with a pair of $q = \pm 2$ charged defects separated by two parallel strings composed of seven achiral cells (green crosses). The string is introduced in a ground state region with cells that possess defined clockwise (orange arrows) or counterclockwise (yellow arrows) chirality. (**b**) Upon application of an external field ($B = 18.5$ mT), the defect pair rapidly annihilates via a sequence of particle flips, restoring the ground state after 18 s (**c**). (**d**) Square ice state in which a single grain boundary of achiral cells separates two incompatible ground state regions. Scale bars, 20 μm for all images. In **e**, the time evolution of the defect line when subjected to the same field condition as in **b** shows no change because ground state recovery requires the high-energy cost of flipping all the spins in one of the two regions. (**f**) Stepwise defect reduction as the field is applied, reflecting the consecutive spin flipping process that leads to the ground state in **b**.

The stability of domain walls between regions of different chirality suggests that one possible mechanism of information storage in the square system can be achieved by arranging the particles in the vertices in such a way as to maximize the number of spin flips required to reach the GS. For example, the triangular pattern of achiral cells shown in Fig. 4a is formed by flipping three of the four spins in a cell. The defects disappear by applying a field of 25 mT, which causes the three particles to shift consecutively, in a manner similar to that presented in Fig. 3b. In contrast, if we set a counterclockwise chiral cell in place of a clockwise chiral cell, surrounded by four achiral cells, fixing the defects requires a simultaneous four-spin reversal because each individual switch leads to a higher energy state. This simultaneous flipping is an energetically higher erasure process, as shown in Fig. 3c. Beyond 25 mT, the pair interactions can become stronger than the lateral confinement, and the colloids have been observed to escape from their gravitational trap in such a way that they rearrange into a triangular lattice. For this reason, we use numerical simulation to determine the threshold field $B_c$ necessary to reset the GS, (details in the Methods). We confirm that the GS is reached by inverting the chirality of the central cell at a field of $B_c = 30$ mT. Cell writing with defined chirality can be used to store digital information in the form of 8 bit ASCII characters. Once written with optical tweezers, the chiral domains are stable below $B_c$. In addition, any low-energy defect, such as those shown in Fig. 4a, easily disappears. More examples of stable and unstable achiral cells are shown in Supplementary Figs 1 and 2, and commented in Supplementary Note 1.

In conclusion, we have engineered artificial colloidal spin ice states in which an external magnetic field fully controls the spin interactions and the collective dynamics leading to the degenerate GS of the system. Unlike the original proposition of a completely optical system[28], our bistable gravitational traps are sculptured in soft-lithographic platforms, can be arranged into periodic lattices and can be designed with diverse geometries and lattice constants. Our geometrically frustrated soft matter system provides a robust approach for probing the effect of disorder by manually introducing defects into the lattice pattern. Disorder in the system can be also created by either leaving traps empty or, as recently proposed[31], by creating sites of double occupation which correspond to pairs of outward pointing spins, not possible with the nanoscale spin ice system. Finally, the strategy presented here can offer guidelines for designing similar experiments on nanoscopic systems or probing the stability of spin arrangements to record information in magnetic data storage devices[32].

## Methods

**Fabrication of the soft-lithographic platform.** The pattern was written via direct write laser lithography (DWL 66, Heidelberg Instruments Mikrotechnik GmbH) on a 5-inch Cr mask with a $\lambda = 405$ nm diode laser at a 5.7 mm$^2$ min$^{-1}$ writing speed. As shown in Supplementary Figs 3 and 4, the small hill in the centre of each trap was obtained by drawing a small constriction in the middle of the elliptical well. These structures were exposed on a ~100-μm-thick coverglass by a 2.8-μm-thick layer of a positive photoresist AZ-1512HS (Microchem, Newton, MA) deposited by spin coating (Spinner Ws-650Sz, Laurell) performed at 1,000 r.p.m. for 30 s. After the deposition, the photoresist was irradiated for 3.5 s with ultraviolet light at a power of 21 mW cm$^{-2}$ (UV-NIL, SUSS Microtech). Later the exposed regions were eliminated by submerging the plate in AZ726MF developer solution for 7 s. Some representative images of the resulting structures are shown in the Supplementary Figs 3–5. The substrate was finally covered with a thin layer of polysodium 4-styrene sulfonate by using the layer-by-layer adsorption technique. More details are given in the Supplementary Note 2.

**Magneto-optical set-up and experiments.** The experimental set-up allows applying simultaneously and independently magnetic and optical forces. It is composed by an inverted homemade optical microscope equipped with a white light illumination LED (MCWHL5 from Thorlabs), a charge-coupled device (Basler A311f) and custom-made coil perpendicular to the sample cell such that the main axis points along the z-direction. The coil was connected to a programmable power supply (KEPCO BOP-20 10M), which is remotely controlled along with the image acquisition and recording with a custom-made LabVIEW programme. The photoresist is sensitive to ultraviolet light, so the white light of the LED is filtered with a long pass filter with a cutoff at 500 nm (FEL0500 Thorlabs). Optical tweezers are realized by tightly focusing a $\lambda = 975$ nm, $P = 330$ mW, Butterfly Laser Diode (Thorlabs) with a 100× Achromatic microscope objective (Nikon, numerical aperture = 1.2) which is also used for observation purpose. During the experiments, a solution is prepared with 3.5 µl of polystyrene paramagnetic particles (PS-MAG-S2874Microparticles GmbH) with 10 ml of high-deionized water (MilliQ system, Millipore). A drop is placed on the soft lithography structures and after few minutes, the particles sediment due to density mismatch until they are suspended above the substrate due to the electrostatic repulsive interactions with the negative charged surface. We use the optical tweezers to fill all the bistable traps with exactly one particle, removing excess or aggregated colloids. After the initial setting, the particles are allowed equilibrate in their wells for ~2 min before applying the external field. We place a total of 84 particles in the square lattice and 64 in the honeycomb lattice in an experimental field of view of $325 \times 222$ µm$^2$. The effect of the boundary on the experimental system is discussed in Supplementary Note 3 and shown in Supplementary Fig. 6.

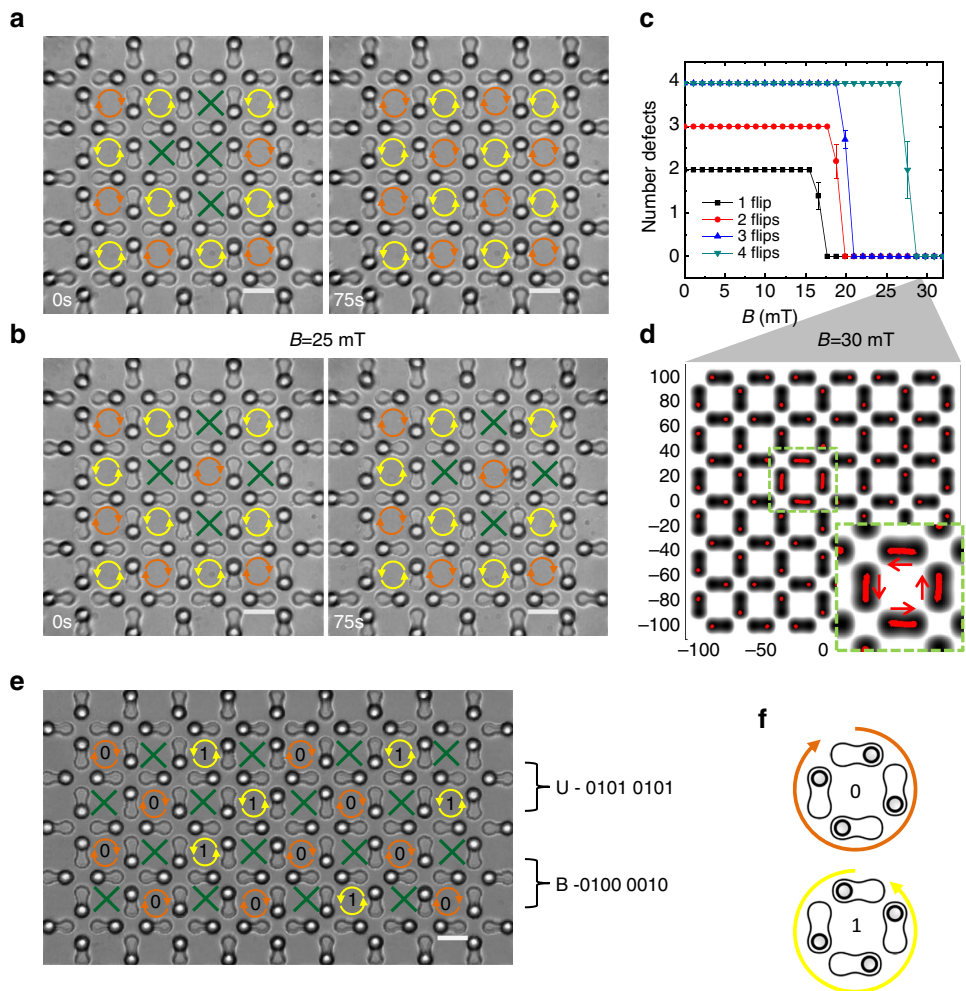

**Figure 4 | Defect stability and binary writing procedure.** (**a,b**) Microscope images showing the evolution with time of three (**a**) and four (**b**) achiral cells in a square lattice when subjected to a static field of $B = 25$ mT. Flipping three spins is necessary in **a** to reach the ground state, while in **b** the defect arrangement requires the inversion of the chirality of the central cell, which is energetically more expensive. (**c**) Reduction of defects versus applied field obtained from numerical simulation. At a field of $B = 30$ mT, (**d**) the simulation shows that it is possible to drive the system to its ground state by the simultaneous flipping of all four particles, inverting the cell's chirality. Inset shows an enlargement of the central cell. (**e**) Example showing the word 'UB' written in a square ice using the binary ASCII representation (8 bit). (**f**) Schematic showing how it is assigned a value of 0 (1) to a clockwise (counterclockwise) chiral cell. Bits in **e** are read from left to right and from top to bottom. Scale bars, 20 µm for all images.

**Brownian dynamics simulation.** We perform two-dimensional Brownian dynamics with periodic boundary conditions containing $N$ particles arranged into an ensemble of double-well traps. A particle $i$ at position $\vec{r}_i \equiv (x_i, y_i)$ obeys the set of overdamped equations:

$$\begin{cases} \eta \dot{x} = \vec{F}_{tot} \cdot \hat{e}_x + \xi_x(t) \\ \eta \dot{y} = \vec{F}_{tot} \cdot \hat{e}_y + \xi_y(t) \end{cases} \quad (1)$$

where $\eta$ is the viscous friction and $\vec{F}_{tot}$ is the sum of external forces acting on the particle, composed by three terms $\vec{F}_{ext} = \vec{F}_g + \vec{F}_N + \vec{F}_M$. Here $\vec{F}_g$ is the gravitational force, $\vec{F}_N$ the normal force exerted by double-well confinement and $\vec{F}_M$ the magnetic interaction between particles. The gravitational force is given by $\vec{F}_g = gV\Delta\rho$, where $g$ is the gravitational acceleration, $V$ is the volume of the particles and $\Delta\rho$ is the density mismatch. We assume the following shape for the double-well potential,

$$z(\delta\vec{r}) = h\left(\frac{2}{d}\right)^4 \left((\delta\vec{r} \cdot \hat{e}_\parallel)^2 - \left(\frac{d}{2}\right)^2\right)^2 + k(\delta\vec{r} \cdot \hat{e}_\perp)^2 \quad (2)$$

where $\delta\vec{r}$ the displacement vector from the centre of the trap, $d$ the distance between the two stable positions, $h$ the height of the central hill and $k$ is the transverse width of the trap. The two unit vectors $\hat{e}_\parallel$ and $\hat{e}_\perp$ define the orientation of the trap; $\hat{e}_\parallel$ is the vector that joins the two stable positions and $\hat{e}_\perp$ is the transverse axis. The corresponding trap is shown in Supplementary Fig. 7a. Assuming the walls have a small inclination angle, the normal force can be calculated as $\vec{F}_N = |\vec{F}_g|\nabla_t z$, which gives:

$$\vec{F}_N = |\vec{F}_g|\left[h\left(\frac{d}{2}\right)^4\left(4\delta r_\parallel^3 - \delta r_\parallel d^2\right)\hat{e}_\parallel + 2k\delta r_\perp \hat{e}_\perp\right] \quad (3)$$

where $\delta r_\parallel = \delta\vec{r} \cdot \hat{e}_\parallel$ and $\delta r_\perp = \delta\vec{r} \cdot \hat{e}_\perp$. The magnetic interaction is calculated assuming every particle has a magnetic moment $\vec{m} = \frac{|\vec{B}|\chi V}{\mu_0}\hat{z}$. Here $|\vec{B}|$ is the amplitude of the magnetic field, $\chi$ is the magnetic volume susceptibility, $\mu_0$ is the permeability of the medium and $V$ is the particle volume. The total magnetic force exerted on one particle is then given by: $\vec{F}_{M_i} = \sum_j \frac{3\mu_0}{2\pi|\vec{r}_{ij}|^4}|\vec{m}|^2 \vec{r}_{ij}$ where $\vec{r}_{ij}$ is the vector that goes from particle $i$ to particle $j$. Finally $\xi(t)$ in Equation (1) is a Gaussian white noise with zero mean, and a correlation function:

$\langle \xi(t)\xi(t') \rangle = 2\eta k_B T \delta(t-t')$. We numerically integrate Equation (1) using a finite time step of 0.01 s and substituting experimental parameters for most quantities. However, the height of the central hill $h$ connecting the two circular traps in the double wells is modified to match the experimental data. Small discrepancies can arise in the hill elevations in photolithographic platforms realized during different fabrication process. First, to validate our theoretical model, we perform initial simulation to match the displacement observed between isolated pair of particles placed within two double well-oriented in a square and honeycomb lattice. The good comparison between experimental data (scattered points) and simulation results (continuous lines) is shown in Supplementary Fig. 7b. The step-like behaviour of the pair distance observed for the blue and green curves is due to the barrier overcoming of one particle. This process is energetically more expensive for isolated particles, reflecting that in the colloidal spin ice system the particle arrangement is a true collective effect rather than resulting from local energy minimization.

## Acknowledgements

This work was supported by the European ResearchCouncil through Starting Grant No. 335040. P.T. acknowledges support from 'Ramon y Cajal' Programme No.RYC-2011-07605 and from Mineco (Grant Mo. FIS2013-41144-P). We thank T.M. Fischer and J. Ortín for fruitful discussions. R. Albalat is acknowledged for help with the AFM measurements and J. Löhr for the profilometer measurements.

## Author contributions

P.T. conceived and supervised the research. A.O.-A. performed the experiments. P.T. and A.O.-A. analysed the data. Both authors wrote the manuscript and discussed all the implications.

## Additional information

**Supplementary Information** accompanies this paper at http://www.nature.com/naturecommunications

**Competing financial interests:** The authors declare no competing financial interests.

**Reprints and permission** information is available online at http://npg.nature.com/reprintsandpermissions/

**How to cite this article**: Ortiz-Ambriz, A. & Tierno, P. Engineering of frustration in colloidal artificial ices realized on microfeatured grooved lattices. *Nat. Commun.* 7:10575 doi: 10.1038/ncomms10575 (2016).